\documentclass[preprint]{aastex}

\def \frmppmhr {631 }
\def \frsppmhr {205 }
\def \frtmppmhr {534 }
\def \frtsppmhr {204 }
\def \fpmppmhr {483 }
\def \fpsppmhr {161 }

\def \mrmppmhr {2143 }
\def \mrsppmhr {1260 }

\def \mptmppmhr {959 }
\def \mptsppmhr {489 }

\def \mrmppm10 {5273 }
\def \mrsppm10 {3086 }
\def \mrtmppm10 {3945 }
\def \mrtsppm10 {2105 }
\def \mpmppm10 {3584 }
\def \mpsppm10 {1941 }
\def \mptmppm10 {2350 }
\def \mptsppm10 {1200 }

\def \hdppmhr {415 }
\def \hdppmff {611 }

\begin{document}
\title{Precision of a Low-Cost InGaAs Detector for Near Infrared Photometry}

\author{Peter W. Sullivan\altaffilmark{1}, Bryce Croll\altaffilmark{1,2}, Robert A. Simcoe\altaffilmark{1}}
\altaffiltext{1}{MIT-Kavli Institute for Astrophysics and Space Research}
\altaffiltext{2}{NASA Carl Sagan Fellow}

\begin{abstract}
We have designed, constructed, and tested an InGaAs near-infrared camera to explore whether low-cost detectors can make small ($\leq$1 m) telescopes capable of precise ($<$1 mmag) infrared photometry of relatively bright targets. The camera is constructed around the 640x512 pixel APS640C sensor built by FLIR Electro-Optical Components. We designed custom analog-to-digital electronics for maximum stability and minimum noise. The InGaAs dark current halves with every 7$^\circ$C of cooling, and we reduce it to 840 $e$-/s/pixel (with a pixel-to-pixel variation of $\pm$ 200 $e$-/s/pixel) by cooling the array to -20$^\circ$C. Beyond this point, glow from the readout dominates. The single-sample read noise of 149 $e-$ is reduced to 54 $e-$ through up-the-ramp sampling. Laboratory testing with a star field generated by a lenslet array shows that 2-star differential photometry is possible to a precision of \frmppmhr $\pm$\frsppmhr ppm (0.68 mmag) hr$^{-1/2}$ at a flux of 2.4$\times 10^4$e-/s. Employing three comparison stars and de-correlating reference signals further improves the precision to \fpmppmhr $\pm$\fpsppmhr ppm (0.52 mmag) hr$^{-1/2}$. Photometric observations of HD80606 and HD80607 ($J$=7.7 and 7.8) in the $Y$ band shows that differential photometry to a precision of \hdppmhr ppm (0.45 mmag) hr$^{-1/2}$ is achieved with an effective telescope aperture of 0.25 m. Next-generation InGaAs detectors should indeed enable Poisson-limited photometry of brighter dwarfs with particular advantage for late-M and L types. In addition, one might acquire near-infrared photometry simultaneously with optical photometry or radial velocity measurements to maximize the return of exoplanet searches with small telescopes.
\end{abstract}

\section{Introduction}

Precise photometry at near-infrared wavelengths remains much more difficult than at optical wavelengths. Atmospheric emission lines are stronger, and their time-variability \citep[e.g.][]{hecht} makes background estimation more critical. Molecular absorption in the atmosphere limits the bands in which one can observe and also varies with time. Perhaps most importantly, observations longer than the silicon cutoff at $\lambda \sim 1\mu$m require detector materials with smaller bandgap energies. This gives near-infrared detectors higher intrinsic dark current than their silicon counterparts for a given temperature. Furthermore, charge-coupled devices (CCDs) are not traditionally fabricated from these materials, so the photosensitive material must instead be hybridized onto a silicon CMOS readout integrated circuit (ROIC). Because charge-to-voltage conversion takes place within each pixel rather than in external electronics, read noise and non-uniformity are higher in CMOS detectors over CCDs. 

Even so, high-performance CMOS infrared detectors, like the Teledyne HAWAII-2RG (H2RG) made from HgCdTe \citep{beletic08}, have revolutionized infrared astronomy over the past decade. The high cost of these detectors and the cryogenic instrumentation built around them limits their deployment to the largest telescopes, however. Indium Gallium Arsenide (InGaAs) is another direct-bandgap semiconductor, but it carries less than half the cost per pixel since it is produced in higher volume than HgCdTe. InGaAs also finds uses in the telecommunications industry, but like HgCdTe, investment from the U.S. Department of Defense is largely responsible for improving the quality of domestically-produced InGaAs imaging arrays. Recent models have attained the pixel counts and performance levels that ought to make them appropriate for infrared imagers on small telescopes that are similarly inexpensive to build and operate. 

\begin{figure}
\epsscale{1.0}
\plotone{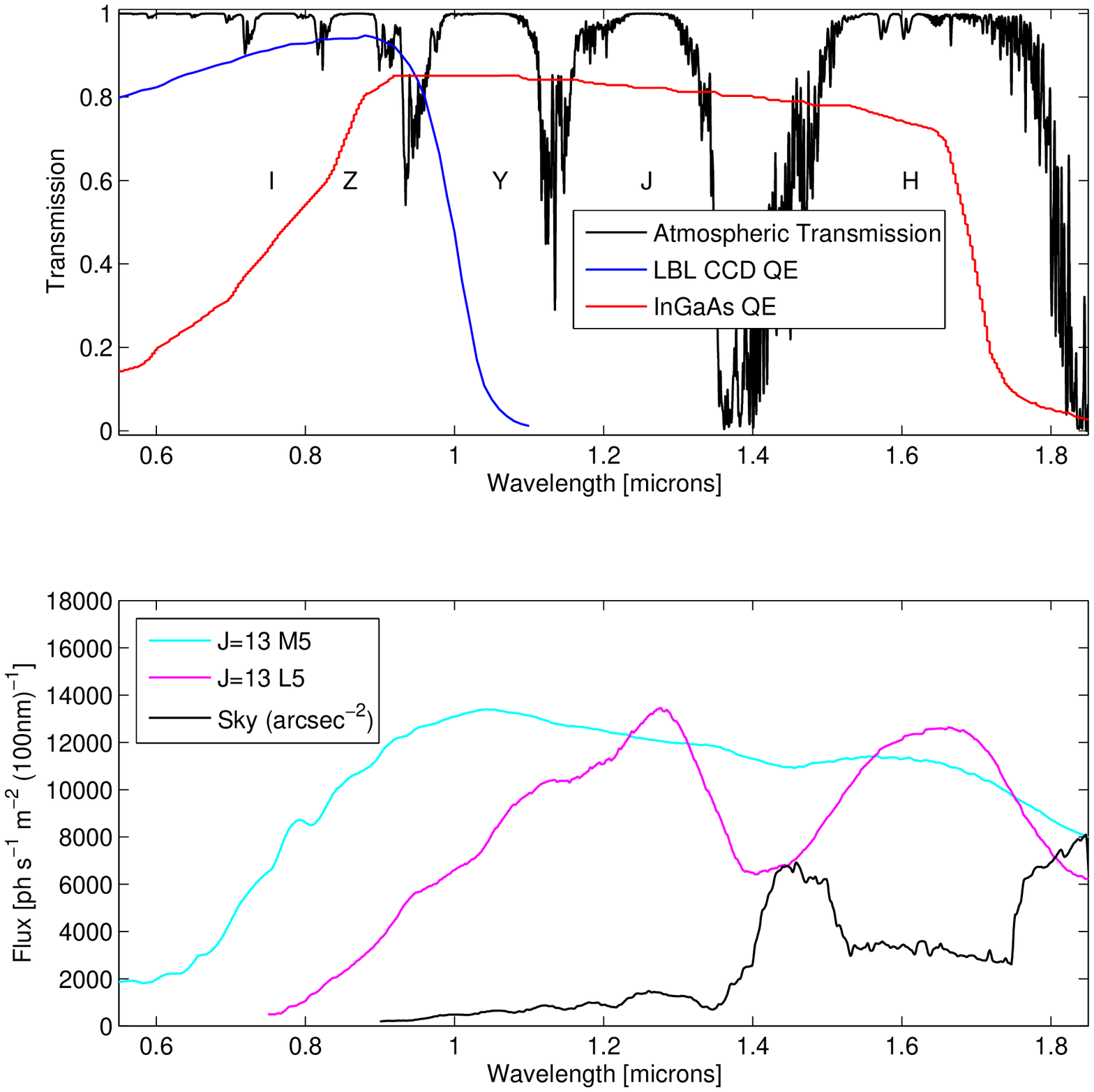} 
\caption{\emph{Top:} Atmospheric transmission \citep[black,][]{linepak} and representative quantum efficiencies of fully-depleted silicon \citep[blue,][]{lblqe} and substrate-thinned InGaAs (red). \emph{Bottom:} A template M5 dwarf spectrum \citep[cyan,][]{uvklib} and L5 dwarf spectrum \citep[magenta,][]{burg04} normalized to $J$=13 as well as atmospheric emission \citep[black,][]{sullisky} convolved with a 100 nm instrumental bandpass. While InGaAs has high quantum efficiency from 0.9 to 1.7$\mu$m, these curves suggest that the $Y$ and/or $J$ bands are optimal for conducting transit photometry of late-type dwarfs.}
\label{bandpass}
\end{figure}

Unlike HgCdTe, which is continuously tunable in its cutoff wavelength, InGaAs is only lattice-matched to its substrate when tuned to 1.7 $\mu$m. This cutoff excludes the $K$ band, but it also makes the detector less sensitive to local thermal emission and renders cryogenic cold shielding unnecessary. This way, an InGaAs detector can be cooled with a Cryotiger (or similar closed-loop cold head) in the same fashion as research-grade CCDs. InGaAs sensor arrays have previously been tested on the HAWAII-1RG ROIC, and the dark current at 140K is very similar to that of HgCdTe tuned to 1.7 $\mu$m \citep{snap07}. Standard, off-the-shelf ROICs optimized for snapshot imaging carry higher read noise than the HAWAII ROICs, but they have higher production volume and yield (due to their smaller size), which reduces cost. The HAWAII family's larger format of (2 - 4 K)$^2$ versus the (0.5 - 1 K)$^2$ of standard ROICs may not be necessary for the image planes of smaller telescopes, and the radiation tolerance and space qualification of the H2RG add unnecessary cost for ground-based telescopes. In the least-expensive implementation, an InGaAs detector operating with a standard ROIC, thermoelectric cooling, and room-temperature electronics can still deliver broadband imaging across the $Y$, $J$, $H$ bands for bright targets. As with HgCdTe, substrate removal extends the responsivity through the $Z$ and $I$ bands. The quantum efficiency curve of InGaAs is plotted in Figure \ref{bandpass}. 

Figure \ref{bandpass} also shows how the spectral response of InGaAs can capture much of the flux from M5 to L5 dwarfs. In contrast, the quantum efficiency of a CCD (which has been optimized for red response by using a thick, fully-depleted detector) still rolls off blueward of the M5 spectral peak. M dwarfs are attractive targets for transit surveys such as the MEarth project \citep{mearth} since Earth-sized planets show larger transit depths in front of these smaller stars and orbital periods in the habitable zone are much shorter. These advantages are even stronger for brown dwarfs, although a transiting circum-brown dwarf planet has yet to be discovered. Surveying the entire sky for new exoplanets or studying a few objects with long-term observations requires dedicated facilities. Such programs can generally be executed with modest telescopes, so the cost of the detector must also be low to make these facilities economical. This is particularly true if multiple telescopes are employed. Even for large telescopes, which are less sensitive to the price of individual detectors, the low cost of InGaAs should be attractive if many detectors are required to sample a wide field at high spatial resolution or if several bandpasses must be observed simultaneously.

The ability to test different filter bandpasses, observational strategies, and signal processing on small, inexpensive telescopes would also help inform observational techniques on larger, more expensive telescopes. Several groups have used the HgCdTe imagers on large telescopes for exoplanet observations in the $J$, $H$, and $K$ bands. However, unknown systematics still drive the precision of these observations well above the photon-counting (Poisson) limit both in photometry \citep[e.g.][]{croll10} and spectroscopy \citep[e.g.][]{bean2011}. 

In either case, low-cost detectors can make it economical to add an infrared capability to modest telescopes. InGaAs arrays hybridized to standard ROICs have not yet been characterized for time-series photometry, but we will examine one such detector here. Section \ref{design} describes the construction of an InGaAs camera and its characterization in the laboratory, Section \ref{lab} describes photometric testing in the laboratory, and Section \ref{sky} describes on-sky testing with the 0.6 m telescope at MIT's Wallace Observatory.

\section{Camera Construction}
\label{design}

\begin{figure}
\epsscale{1.0}
\plottwo{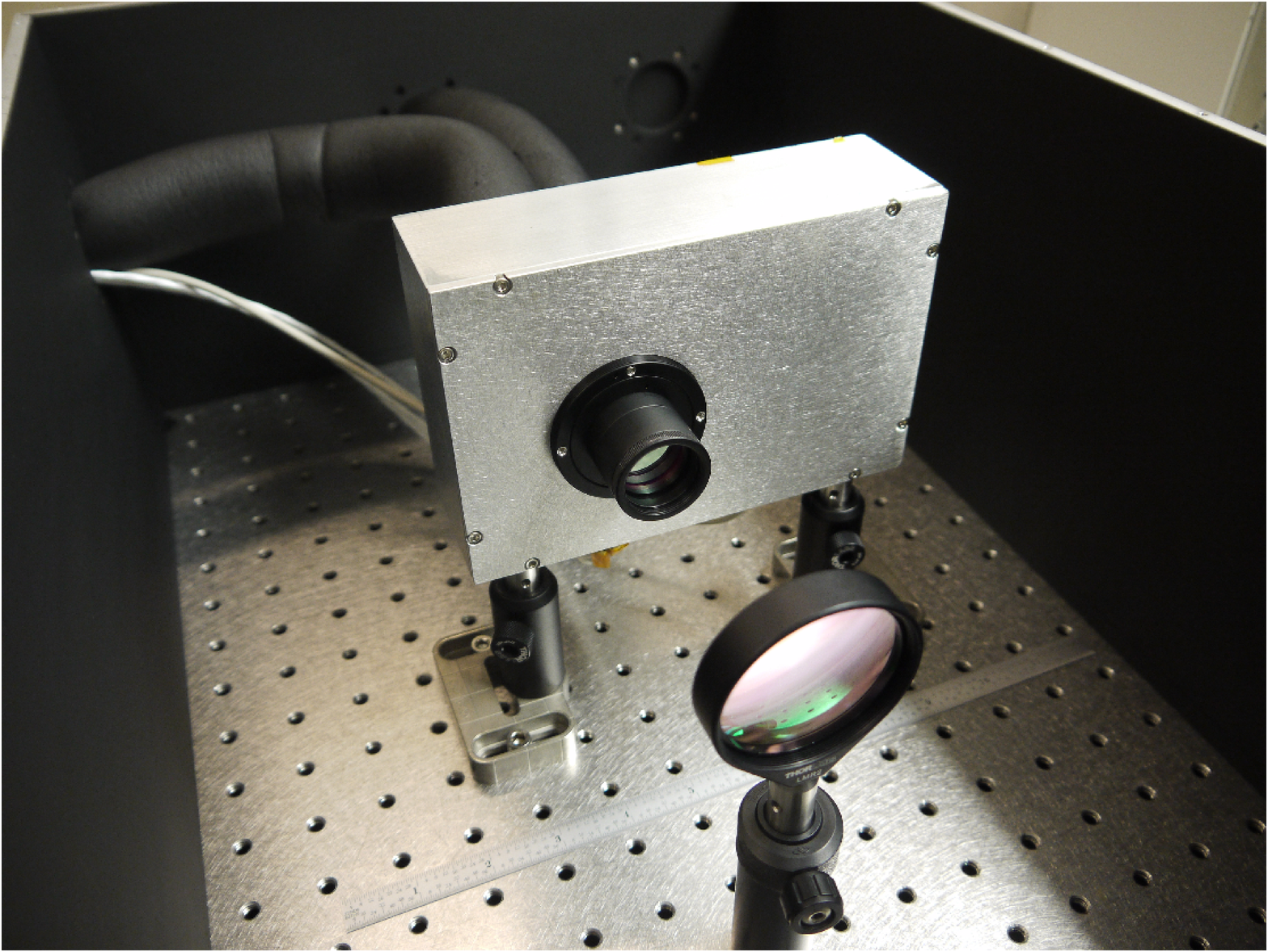}{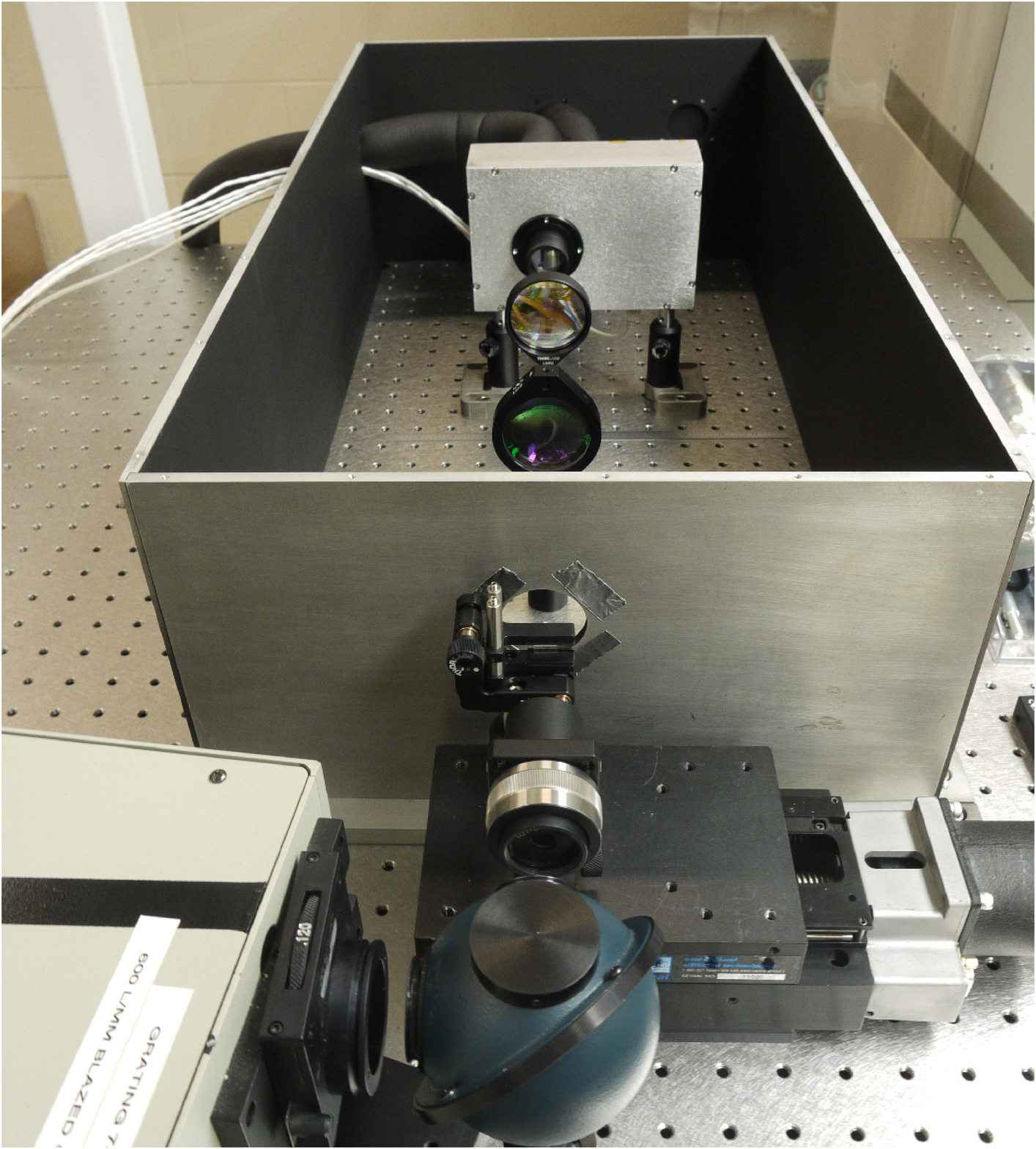}
\caption{\emph{Left:} The assembled camera shown in the laboratory. The chilled water lines are visible behind the camera. \emph{Right:} The optical setup for testing with the lenslet array. From bottom to top, it consists of an integrating sphere (blue), pinhole and collimating lens, microlens array, and a two-lens relay inside a blackened housing.}
\label{pix}
\end{figure}

We first examined fully-assembled cameras available off-the-shelf, specifically looking for units with large cooling capacity to minimize dark current; low-noise, non-destructive readout; high-resolution analog-to-digital conversion; and sufficiently low cost to allow fielding the sensor on multiple telescopes. Such a camera was not available on the commercial market, so we proceeded with constructing our own camera with these requirements in mind. It is built around the APS640C InGaAs image sensor supplied by FLIR Electro-Optical Components with 640x512 pixels spaced at 25 $\mu$m \citep{aerius}. To reduce cost, we acquired a ``B''-grade sensor which is allowed more inoperable pixels (especially in clusters) than the ``A'' grade, but we took care to exclude the inoperable pixels from our observations and analyses. The assembled camera is shown in Figure \ref{pix}.

\subsection{Thermal Considerations}
Domestically-produced InGaAs has reached dark current densities of 1-2 nA/cm$^2$ at room temperature. The dark current halves with every 7$^{\circ}$C of cooling \citep{aerius}, so we sought to minimize the operating temperature of the detector. The InGaAs detector is hybridized to a silicon read-out integrated circuit (ROIC) that is housed in an evacuated package with a thermo-electric cooler (TEC) sandwiched between the ROIC and the housing. A standard laboratory closed-loop water chiller holds the warm side of the TEC at 10$^{\circ}$C, which is above the dewpoint during favorable observing conditions. The TEC drives a $\Delta$T of -40$^{\circ}$C, allowing the detector to operate down to -30$^{\circ}$C. Although water cooling complicates the operation of the detector, the heat exchanger can be moved far from the telescope whereas a fan mounted to the camera might introduce electrical or mechanical noise.

\subsection{Analog-to-Digital Electronics}
The higher read noise that CMOS detectors exhibit over CCDs can be mitigated by taking multiple samples during an exposure if the CMOS device supports non-destructive readout. This enables one to sample at the beginning and end of an exposure, known as digital correlated double-sampling (dCDS) or Fowler pairs \citep{fowler}, as well as sampling throughout the exposure and fitting a line the accumulated charge over time, known as up-the-ramp sampling \citep{sutr}. In order to allow $\sim$10 frames to be taken up-the-ramp during exposures lasting several seconds, we digitize the analog output of the detector through one channel at 1 MHz. While using multiple detector outputs or a higher digitization rate could increase the frame rate beyond the 2.7 Hz we achieved, this design makes any instabilities in the signal chain common to all pixels, simplifies the analog design, and eases the power consumption of the ROIC and its cooling needs. All pixels across the array are sampled simultaneously since each pixel has a sample-and-hold. In other words, it has a snapshot electronic shutter and not a rolling electronic shutter like the H2RG.

We connected the ROIC's analog output to a 16-bit analog-to-digital converter (ADC) via an op-amp buffer. A 12-bit digital-to-analog converter (DAC) drives the bias inputs on the ROIC and reference voltages in the ADC chain. Both the ADC and the DAC are tied to the same low-noise, low-drift (3 ppm/$^{\circ}$C) voltage reference, which ensures the electronic stability of the camera. Relatively few other parts are needed to operate the sensor.

\subsection{Digital Design}
Clocking of the detector and communication between the data converters and the host computer is handled by a Virtex Spartan 6 field-programmable gate array (FPGA). It is housed on a daughter circuit board made by Opal-Kelly that provides access to the FPGA's input/output pins as well as a Universal Serial Bus (USB) interface; a library supplied by Opal-Kelly supports data transfers between FPGA registers and data structures in the host computer's software. The FPGA is powered by the computer via the USB cable. While USB provides ample throughput, it also introduces noise from the host computer's power supplies. To keep this noise from entering the analog portion of the circuit board, and to break the ground loop between the computer and camera chassis, digital isolators are placed on the lines connecting the ADCs and DAC to the FPGA. 

The clocking patterns which drive the image sensor and ADCs are designed to be as regular as possible to minimize thermal transients. In fact, the ADCs continuously convert the detector output whenever the camera is powered on; when pixels are read out, the only change is that the ADC values are piped through the USB interface rather than terminating in the FPGA.

\subsection{Detector Characterization}
\begin{figure}
\epsscale{0.6}
\plotone{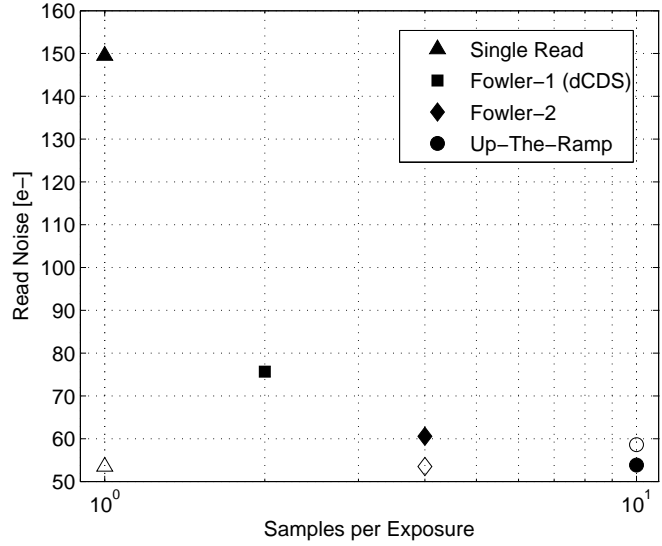}
\caption{Multiple sampling during an exposure significantly reduces the effective read noise. Filled markers indicate measured values, and open markers indicate values predicted from \citet{offenberg} based on the Fowler-1 read noise. The single-sample read noise lies far above the predicted value since reset ($k_BTC$) noise is present. The Fowler-4 and sample-up-the-ramp noise performance is consistent with expectation.}
\label{rnoise}
\end{figure}

We measured the gain, read noise, and dark current from a series of dark and flat-field frames (illuminated with a lamp). By comparing the shot noise of illuminated and dark frames \citep[following][]{janesick}, we measure the gain to be 0.6 $e$-/ADU. The read noise was obtained from extrapolation to zero signal, which is 149 $e$- if one sample is taken per frame. This is larger than the 70 $e$- specified by the manufacturer, but our longer exposure time might amplify any $1/f$ noise in the InGaAs photodiodes or the ROIC. Taking multiple samples per frame (either in Fowler pairs or sampling up-the-ramp) cancels out the reset noise on the integration capacitor and dramatically reduces the read noise, as shown in Figure \ref{rnoise}. Ramps of 10 samples minimize the effective read noise to 54 $e$-, which is consistent with prediction from the Fowler-1 read noise value.

\begin{figure}
\epsscale{1.0}
\plotone{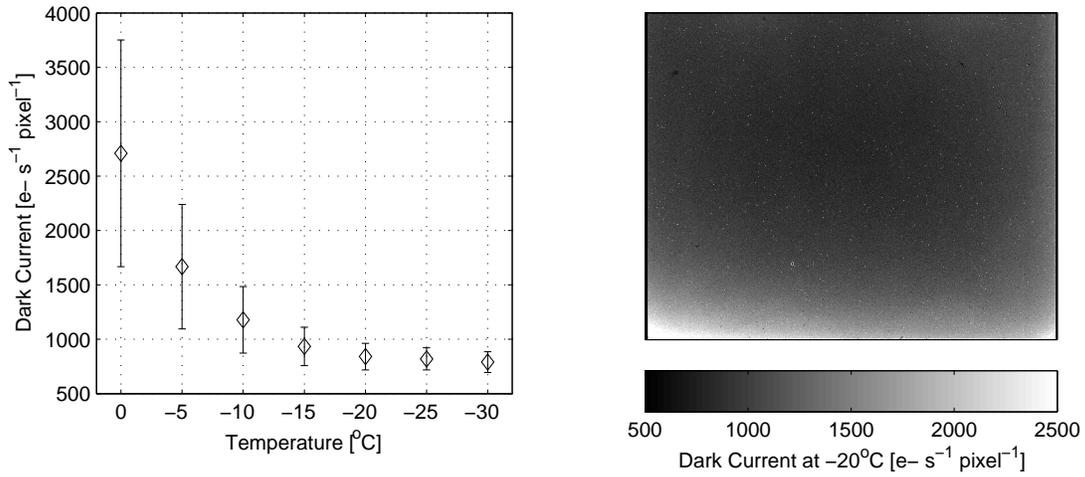}
\caption{\emph{Left:} Dark current versus detector temperature. The error bars indicate the 1$\sigma$ pixel-to-pixel spread in the measured values. No further reduction in dark current is possible past -20$^{\circ}$C. \emph{Right:} The spatial pattern of the dark current across the array. Because it is stronger near the edges of the detector, we attribute the excess dark current to ROIC glow.}
\label{darkcurrent}
\end{figure}

Dark current was also measured in this setup over temperature and is plotted in Figure \ref{darkcurrent}. Although we expect the dark current to halve for every $\sim 7^{\circ}$C of cooling, we find that it plateaus near the -20$^{\circ}$C value of 840 $e$-/pixel/s. The dark current is even higher in the corners and edges of the array, shown in Figure \ref{darkcurrent}, which indicates that recombination glow from the ROIC is the source. FLIR has acknowledged this issue and has stated (in personal communication) that upcoming designs will reduce ROIC glow. As we will demonstrate, dark current is the dominant source of noise in our photometry on the sky.

Burst noise (or popcorn noise), which is thought to be caused by the filling and emptying of trapping sites in a semiconductor, adds noise on top of that expected from read noise and dark current alone \citep{burst}. Pixels with burst noise will suddenly gain or lose an offset, giving their timeseries an excess kurtosis due to the non-Gaussian distribution of the noise. Approximately 0.1$\%$ of pixels display burst noise, which is similar to that of the H2RG \citep{me11}. 

Under flat-field illumination, pixel-to-pixel non-uniformity is measured at approximately 4.0$\%$ among the operable pixels.

\section{Lenslet Photometry}
\label{lab}
We obtained a long photometric time series with a simulated star field in the laboratory. In a fashion similar to that of \citet{clan12}, a microlens array generates the ensemble of stars. Figure \ref{pix} shows our setup, where a 50 $\mu$m pinhole is illuminated with a lamp and integrating sphere and the microlens array re-images the pinhole into an 8x10 grid of simulated stars with a spacing of 1.015 mm. This pattern is then re-imaged onto the detector with slight magnification, which is shown in Figure \ref{ulens}. The pinhole and microlens relay is attached to a linear stage to allow us to move the image across the detector in the horizontal direction. To flat-field the detector, the pinhole and microlens array are removed to allow the output of an integrating sphere to be imaged directly. Dark frames were also acquired with the shutter on the lamp closed. The median flux per aperture is 2.36$\times 10^4$e-/s, similar to a J$\approx$9.1 star imaged in the Y band with a 0.25 m effective aperture. The imaged stars have a full-width half-maximum of 3 pixels. Individual exposures lasted 3.7 seconds, and a total of 9600 images were acquired over nearly 11 hours.

In order to address the concern that CMOS arrays suffer from inter-pixel variation that will degrade the photometric precision in the presence of tracking error, we repeated the lenslet experiment while dithering the star field in pseudo-random fashion. The star field was moved by a random amplitude ($\pm$1 or 2 pixels) every 50 frames or 3.4 minutes; 2200 images were acquired in this mode over 2.5 hours.

\subsection{Up-the-ramp Aperture Photometry}
Aperture photometry with CMOS arrays is traditionally performed on the image after the sample-up-the-ramp calculation is performed for each pixel. The value of all pixels within an inner radius are summed; the local background is estimated from the median of an outer annulus of pixels and subtracted from this sum. In our implementation of traditional aperture photometry, the radii are centered on the star with the \texttt{FIND} procedure in IDL. 

One can also sum the aperture counts on each frame comprising the ramp \textit{before} the up-the-ramp calculation is performed. In this case, the ramp is fitted to the background-subtracted measurements of flux falling in the photometric aperture. The traditionally-calculated ramp image is still used to center the photometric apertures on the stars in the frame. While the two approaches may appear to be equivalent, we find there a 2$\%$ reduction in the noise of the photometric timeseries taken in the laboratory but no improvement with photometry on the sky when using up-the-ramp aperture photometry. 

There are two reasons why up-the-ramp aperture photometry should perform better. First, noise common to all pixels on the array is subtracted in each frame comprising the ramp; it is difficult to remove this common-mode noise without knowing \textit{a priori} where the stars lie in the frame. Secondly, atmospheric seeing may distribute the flux in a non-uniform fashion across the PSF during the exposure. The flux accumulated over the whole aperture is therefore a more linear function of time than the flux accumulated by a single pixel. While we do not see a significant performance improvement with this technique, it may prove useful elsewhere, and we still employ it here for the sake of uniformity.

\subsection{Measurement of Precision}
We measure the precision in the photometric time series as the relative standard deviation as a function of co-averaging time. Hence, we divide the timeseries $f_i$ into $M$ blocks of duration $\tau$, calculate the mean within the blocks, and find the standard deviation of the block means. If each of the block means are denoted $\bar{f}_j$, we have
\begin{equation}
\bar{f}_j = \frac{1}{N}\sum_{i=Nj}^{N(j+1)}{f_i}
\end{equation}
where $N=\tau/\Delta\tau$ gives the number of samples for a sample period $\Delta\tau$. The relative standard deviation over the block means can then be written
\begin{equation}
\sigma (\tau)  = \frac{1}{\bar{f}}\sqrt{\frac{\sum_j^M{\left(\bar{f}_j-\bar{f}\right)^2}}{M-1}}
\end{equation}
where $\bar{f}$ denotes the mean of $f_i$ over the whole timeseries. Care is taken to ensure that the same time samples are always used in the calculation by using block lengths $N$ that divide into the total number of samples without remainder. In other words, the product $MN$ is always equal to the total length of the time series. For uncorrelated noise, we expect $\sigma(\tau) \propto \tau^{-1/2}$.


\subsection{Reference Signal Subtraction}
In addition to the photometry, we also obtain a set of reference signals during the data acquisition. Before and after each row of pixels is read out from the detector, the bias level is sampled in a similar fashion to the ``overscan'' region of a CCD. Rows of blind pixels on the top and bottom of the detector are also read out at the beginning and end of each frame, which should track changes in the column amplifiers. Furthermore, we record the centroid position of the photometric aperture to correct for residual flat-fielding errors when the stars and apertures are moving. 

De-correlation of the reference signals and centroid position from the photometric timeseries has proved effective with Teledyne H2RG detectors (e.g., \citet{me11} and \citet{clan12}), and we apply the same basic approach here. A robust linear regression determines the fit coefficients $c_h$ between the measured photometric signal $f_i$ and the reference signals $R_{hi}$: 
\begin{equation}
f_i = f^0_i + c_1R_{1i} + c_2R_{2i} + \ldots
\end{equation}
The de-correlation estimates the true photometric timeseries $f^0_i$ by subtracting the reference signals and centroid position from the measured photometric signal. The reference signals are cross-correlated with one another, so we construct their principal components prior to de-correlation so that the linear regression yields a unique solution. However, we do not use PCA to reduce the dimensionality of the reference signal set, as was done in the H2RG papers. 

\begin{figure}
\epsscale{1.0}
\plotone{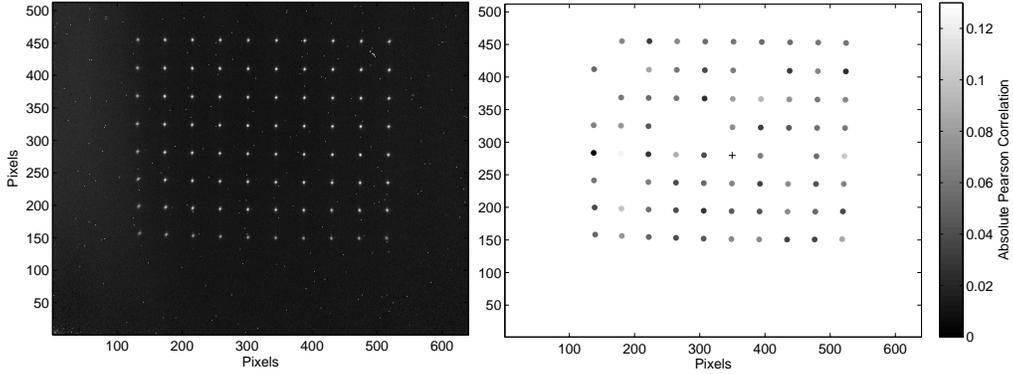}
\caption{\emph{Left:} Full-frame image of stars generated with the lenslet array. \emph{Right:} Absolute value of the Pearson correlation between a simulated star at the center of the field ($+$) and the other simulated stars; there is no clear trend across the array. The diagram omits stars exhibiting burst noise.}
\label{ulens}
\end{figure}

\begin{figure}
\epsscale{1.0}
\plotone{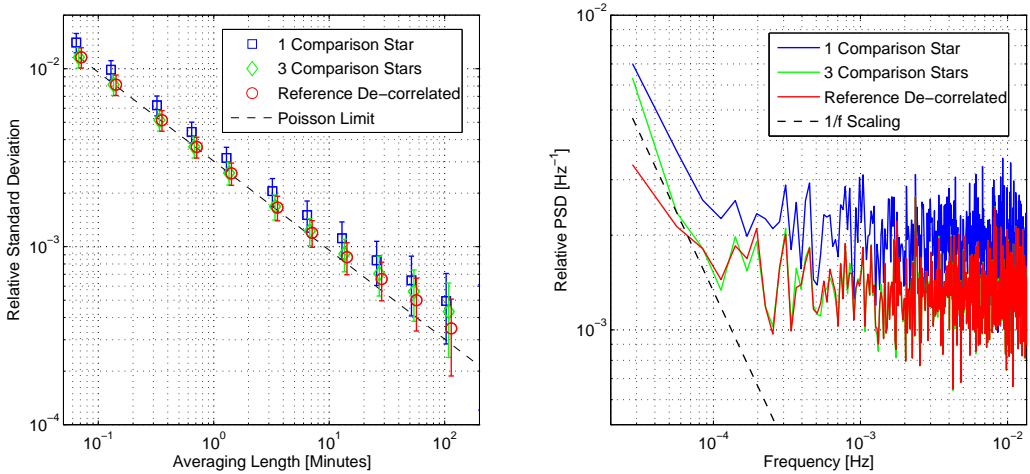}
\caption{\emph{Left:} Precision achieved on an ensemble of 72 simulated stars with differential photometry between one target and one comparison star (blue), three comparison stars (green), and three comparison stars with reference signal de-correlation (red). The $\tau^{-1/2}$ expectation scaled to the leftmost data point (with no averaging) is shown as the dashed line. \emph{Right:} Noise power spectrum for the differential photometry. Adding more comparison stars (green) reduces the white noise floor, and the reference signal de-correlation (red) reduces the $1/f$ component (dashed line).}
\label{fixed}
\end{figure}

\begin{table}
\begin{center}
\caption{Noise budget for a single exposure in laboratory and on-sky testing.\label{noise}\newline}
\begin{tabular}{r|c|c}
\tableline
Noise Source & Lab Photometry & HD80606 Photometry \\
\tableline
Number of pixels & 79 & 201 \\
Read noise (at 54 $e-$/pixel) [$e$-] & 480 & 766 \\
Dark noise (at 840$e$-/pixel/sec) [$e$-] & 498 & 794 \\
Sky noise [$e$-]  & 0 & 113 \\
Poisson noise [$e$-] & 295 & 470 \\
Scintillation noise [$e$-]  & 0 & 493\\
\tableline
Quadrature sum of noise sources & 752 & 1301\\
\tableline
$\sqrt{2}$ increase for differential noise [$e$-] & 1063 & 1840 \\
\tableline
Measured Noise [$e$-] & 1172 & 2740 \\
\tableline
Explained Noise & 91$\%$ & 67$\%$ \\
\end{tabular}
\end{center}
\end{table}

\begin{figure}
\epsscale{1.0}
\plotone{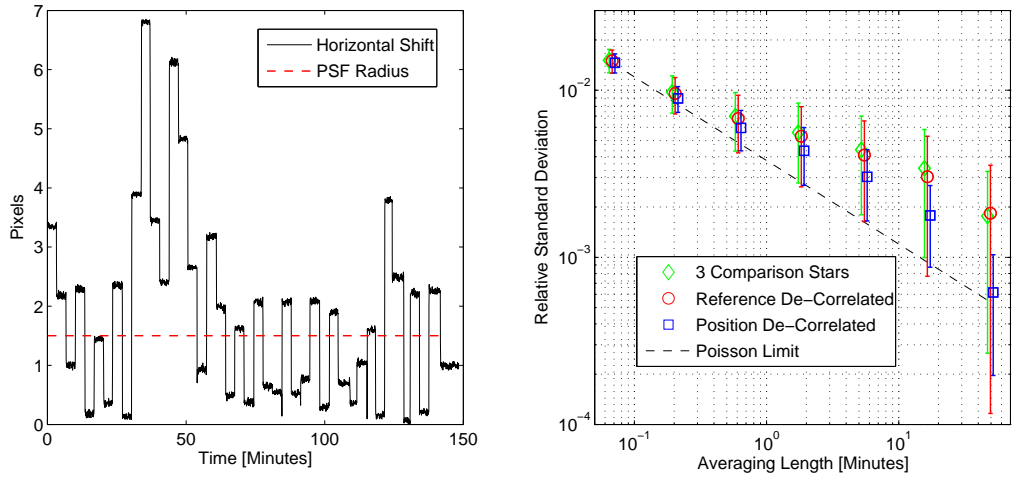}
\caption{\emph{Left:} Dithering of the star pattern in a random fashion to simulate poor telescope tracking. The pseudo-random walk extends beyond the FWHM of the PSF, which is 3 pixels for most of the stars in the ensemble. \emph{Right:} The photometric precision is noticeably worse than with fixed stars, but de-correlating the pixel position from the photometry shows significant improvement. }
\label{move}
\end{figure}

\subsection{Lenslet Results}
\label{res}
With the star pattern fixed, we obtained a time series of 9600 frames in 10.8 hours. Of the 80 simulated stars, 72 produce noise near 1$\%$ per exposure. The other 8 apertures contained at least one pixel exhibiting burst noise, which is consistent with our finding that $\sim$0.1$\%$ pixels show burst noise. Table \ref{noise} shows a breakdown of the calculated sources of noise in the measurement. Read noise, dark current, and Poisson noise alone can account for 91$\%$ of the measured noise. 

Differential photometry is performed for each of the 72 target stars by selecting one comparison star at random from the remaining 71 and finding the ratio of the flux. The comparison stars were chosen at random; as Figure \ref{ulens}b shows, there is no obvious advantage (in terms of a higher Pearson correlation) to selecting comparison stars near the target star. The precision of differential aperture photometry scales as $\tau^{-1/2}$ as expected for uncorrelated noise at \frmppmhr ppm hr$^{-1/2}$ with a standard deviation of $\pm$\frsppmhr across the 72 apertures (Figure \ref{fixed}, left). The median flux per aperture was 2.36$\times 10^4$e-/s. We repeated the analysis with the mean of three comparison stars, which improves the precision to \frtmppmhr $\pm$\frtsppmhr ppm hr$^{-1/2}$. If we de-correlate the reference signals from the photometry, the precision further improves to \fpmppmhr $\pm$\fpsppmhr ppm hr$^{-1/2}$. 

The effects of adding comparison stars and de-correlating the reference signals can also be seen in the frequency domain. In Figure \ref{fixed} (right), both white noise and $1/f$ (red) noise are evident. Averaging over three comparison stars lowers their collective Poisson noise and hence the white noise floor by 30$\%$, but the $1/f$ component remains. The reference signals can reduce the $1/f$ component, but they do not lower the white noise floor at higher frequencies. In all cases, the $1/f$ noise becomes significant at timescales longer than one hour, and approximately 0.5$\%$ of the total noise power is attributed to $1/f$ noise above the white noise floor. 

We also applied the time-domain analysis to the dithered photometry and arrived at the precision shown in Figure \ref{move}. The precision does indeed degrade at the timescales close to the dithering period of 3.4 minutes, and with 3 comparison stars, it only reaches \mrtmppm10 $\pm$\mrtsppm10 ppm in 10 minutes (scaling as \mrmppmhr $\pm$\mrsppmhr ppm hr$^{-1/2}$). De-correlating the reference signals without the $x$ centroid position only improves the 10-minute precision by 9$\%$. Including the $x$ centroid position in addition to the reference signals in the de-correlation shows much greater improvement, yielding a precision of \mptmppm10 $\pm$\mptsppm10 ppm in 10 minutes. This scales as \mptmppmhr $\pm$\mptsppmhr ppm hr$^{-1/2}$, a factor of 2 higher than the case with fixed stars. We feel this test was more stringent than the data actually taken on-sky (see Section \ref{sky}) because the PSF radius was smaller and it was more difficult to flat-field the detector in our laboratory. Even so, the flat-field errors are sufficiently averaged at long ($>$50-minute) timescales to allow the de-correlated signal to return near the Poisson limit. We do not compute the power spectrum for this data set due to its shorter length. 

\begin{figure}
\epsscale{1.0}
\plotone{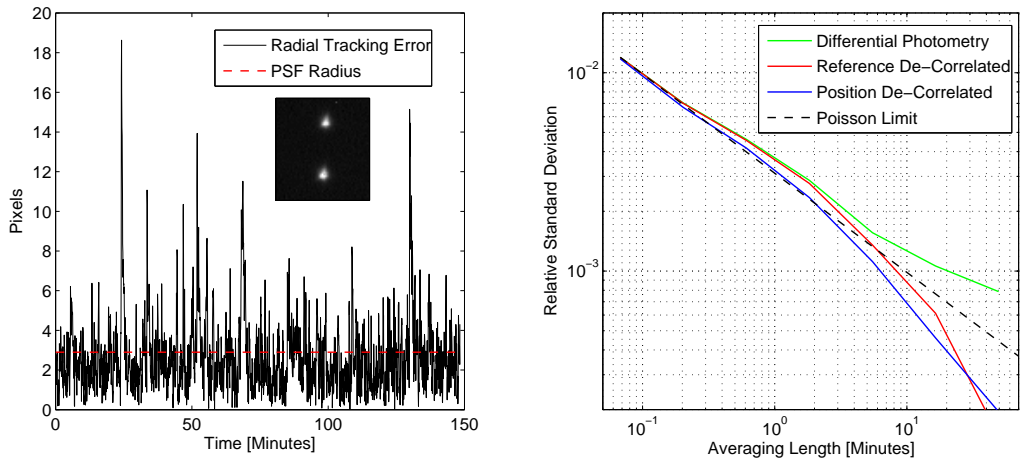}
\caption{\emph{Left:} Telescope tracking errors frequently exceeded the PSF width during observations of HD80606 and HD80607 (inset). \emph{Right:} Precision achieved in differential photometry.}
\label{hd}
\end{figure}

\clearpage

\section{Testing on the Sky}
\label{sky}
\subsection{Observations}
In February 2013, we observed HD80606 and HD80607 with the 0.6 m telescope at MIT's Wallace Astrophysical Observatory. Both stars are spectral type G5 \citep{feltz}, with HD80606 at $J=7.73$ and HD80607 at $J=7.80$ \citep{2mass03}. Although HD80606 is known to host a planet \citep{naef}, the transit \citep{hdtrans} of HD80606b was not occurring during our observation. The observations were carried out in the $Y$ band $\lambda \sim 1.05 \mu$m using a filter from Omega Optical. Guiding commands were derived from the science frames; the tracking errors are plotted in Figure \ref{hd}. Again, we obtained 2200 images in 2.5 hours with an exposure time of 3.7 seconds. Flat-field frames were obtained with a screen in the dome.

\subsection{Sky Results}
Many practical issues make testing on the sky more difficult than in the laboratory. At our site in northern Massachusetts, the seeing was poor (at several arcseconds), requiring photometric apertures with a radius of 8 pixels for our plate scale. Also, we estimate the optical efficiency of the telescope at only 18$\%$ even after accounting for all other known losses, including secondary obstruction and filter transmission. This reduces our effective aperture to 25 cm. Atmospheric scintillation noise is also significant for a telescope diameter of $D$=60 cm and exposure time of $\tau_{exp}$=3.7 seconds; we calculate it at $X$=1.3 airmasses from \citet{young67} and \citet{young93} with the wavelength $\lambda$ scaled from the $V$ to $Y$ band using \citet{dravins97}:
\begin{equation}
\sigma = \left(9\%\right)\left(X\right)^{3/2}\left(D_{cm}\right)^{-2/3}\left(2\tau_{exp}\right)^{-1/2}\left(\frac{\lambda}{0.55 \ \mu m}\right)^{-7/12}\exp\left(-\frac{z}{8 \ \mathrm{km}}\right),
\end{equation}
where $z$ denotes the site altitude in km. Table \ref{noise} shows that, unlike with the laboratory data, we can account for only 67$\%$ of the observed noise by considering the read noise and dark current levels measured in the laboratory, Poisson noise expected from the stellar flux and sky, and scintillation noise. Higher electronic noise at the telescope, underestimated background, and tracking errors could also contribute to the higher observed noise. The laboratory testing showed that gross tracking errors, where the flux centroid strayed from its mean location beyond the PSF half-maximum 56$\%$ of the time, can degrade the precision by a factor of 2. Hence, a degradation by a factor of 1.5 can be accounted by the tracking errors at the telescope, where the flux centroid strayed from the PSF half-maximum 35$\%$ of the time.

Despite these issues, the differential photometry (Figure \ref{hd}) yields a precision down to \hdppmff ppm in 45 minutes for a mean observed flux of 6.7$\times 10^4$ e-/s. Over long timescales, de-correlating the reference signals (with or without the 2-dimensional centroid position) yields precision that is consistent with the Poisson limit of \hdppmhr ppm hr$^{-1/2}$. The actual values of the relative standard deviation fall below the Poisson limit for long averaging times, but with only two stars there is large uncertainty in measuring the precision. Over 1-minute timescales, the precision with the centroid position de-correlated shows 12$\%$ improvement over the precision with only the reference signals de-correlated. This brings the precision into agreement with the Poisson limit, indicating that the effects of tracking and flat-field errors are overcome.


\section{Conclusions}
We have demonstrated that presently-available InGaAs arrays have the necessary stability to perform ground-based near-infrared photometry with precision at the level of 500 ppm hr$^{-1/2}$ or 0.5 mmag hr$^{-1/2}$ at flux levels of 2.4$\times 10^4$ e-/s in the laboratory and 6.4$\times 10^4$ e-/s on the sky. Hence, a 2 R$_{\oplus}$ Super-Earth around a 0.2 R$_{\odot}$, $J$=7.8 M5 dwarf, which gives a transit depth of 0.84$\%$, could be detected to 4$\sigma$ (2100 ppm) in 3 minutes of observation on the Wallace 0.6 m telescope. If one were to use a facility with higher optical efficiency ($\sim$50$\%$ versus 18$\%$) and better seeing (1 arcsecond versus 5), the same detection could be made with a $J\sim$9 host. This is similar to MEarth's detection of GJ1214b, which has a 1.5$\%$ transit depth around a $J$=9.25, M4.5 dwarf \citep{gj1214b}. 

Although CMOS arrays support semiconductors sensitive to wavelengths beyond the red cutoff of silicon, these detectors also have intrinsically higher pixel-to-pixel variation than CCDs from the charge amplifier being placed in each pixel. With 4$\%$ variation in the array tested here, it is crucial to track stars within the PSF if one wants to achieve 500 ppm photometry. Even then, one must perform careful flat-fielding of the array and de-correlate the centroid shifts from the photometry to correct residual errors. Burst noise affects sufficiently few pixels to allow differential photometry of a single target and several comparison stars if the field is pointed to avoid the affected pixels. However, the under-sampled, wide-field surveys currently underway with CCDs would have a high proportion of sources contaminated with burst noise unless data reduction algorithms can be designed to overcome this phenomenon. Dark current from ROIC glow is still the dominant source of noise in the APS640C sensor tested here with 25 $\mu$m pixels. However, upcoming InGaAs arrays with 15 $\mu$m pixels should have less than half the dark current per pixel as dark current scales with the pixel area, and lower operating temperatures will become useful once ROIC glow is reduced.

Besides the photometry of late-type dwarfs, near-infrared photometry may have other applications to exoplanet science with small telescopes. Star spots show reduced contrast at longer wavelengths, and limb darkening is less pronounced, which should improve transit timing measurements. Dichroic beamsplitters can allow simultaneous optical and near-infrared photometry, which can help mitigate systematic errors and reject background binaries due to their color dependence. Optical radial velocity measurements and near-infrared photometry can also be carried out simultaneously on the same telescope; \citet{rv} has shown that simultaneous photometry can help one correct radial velocity deviations from star spots. Thermal emission of hot Jupiters can be measured through secondary eclipse (e.g. \citet{charb05} and \citet{croll10}), and full phase curves of hot Jupiters can be established by acquiring long timeseries. With the improvements that next-generation designs will deliver, InGaAs arrays may maximize the return of exoplanet observations with 0.4-1.0 m or even larger telescopes.


\acknowledgements
We would like to thank the staff of MIT's Wallace Observatory, Tim Brothers and Michael Person, for trusting us with their telescope. Joshua Winn provided useful comments for the manuscript. The MIT CCD Laboratory staff consulted on the design and loaned us electrical tools. Andrew Hood and Falgun Patel at FLIR provided timely technical support for their product. PWS thanks Edward Hugo Darlington at JHU/APL for his mentorship in detectors. Hardware for this project was purchased with the MIT-Kavli Institute Development fund. BC's work was performed under contract with the California Institute of Technology funded by NASA through the Sagan Fellowship Program. The Natural Sciences and Engineering Research Council of Canada supports the research of BC. RAS was backed by the Adam J. Burgasser Chair in Astrophysics.

Facilities: \facility{MIT:WAO:0.61m}

\bibliography{ms}


\end{document}